\documentclass[%
 reprint,
 superscriptaddress,
 amsmath,amssymb,
 aps,
]{revtex4-1}

\usepackage{graphicx}% Include figure files
\usepackage{dcolumn}% Align table columns on decimal point
\usepackage{bm}% bold math
\begin{document}

%\preprint{APS/123-QED}

\title{Simple rules govern the patterns of Arctic sea ice melt ponds}% Force line breaks with \\
\author{Predrag Popovi\'{c}}
 \altaffiliation{All correspondence should be directed to Predrag Popovi\'{c}}%Lines break automatically or can be forced with \\
  \email{ppopovic@uchicago.edu}
  \affiliation{%
 The University of Chicago, The Department of the Geophysical Sciences\\
}%
\author{B. B. Cael}%
\affiliation{%
 Massachusetts Institute of Technology, Department of Earth, Atmospheric and Planetary Sciences\\
}%
\author{Mary Silber}
\affiliation{%
 The University of Chicago, Department of Statistics and Committee on Computational and Applied Mathematics\\
}%
\author{Dorian S. Abbot}
\affiliation{%
 The University of Chicago, Department of Geophysical Sciences\\
}%


%\affiliation{
%\affiliation{
%\affiliation{%


\date{\today}% It is always \today, today,

\begin{abstract}
Climate change, amplified in the far north, has led to rapid sea ice decline in recent years. In the summer, melt ponds form on the surface of Arctic sea ice, significantly lowering the ice reflectivity (albedo) and thereby accelerating ice melt. Pond geometry controls the details of this crucial feedback; however, a reliable model of pond geometry does not currently exist. Here we show that a simple model of voids surrounding randomly sized and placed overlapping circles reproduces the essential features of pond patterns. The only two model parameters, characteristic circle radius and coverage fraction, are chosen by comparing, between the model and the aerial photographs of the ponds, two correlation functions which determine the typical pond size and their connectedness. Using these parameters, the void model robustly reproduces the ponds' area-perimeter and area-abundance relationships over more than 6 orders of magnitude. By analyzing the correlation functions of ponds on several dates, we also find that the pond scale and the connectedness are surprisingly constant across different years and ice types. Moreover, we find that ponds resemble percolation clusters near the percolation threshold. These results demonstrate that the geometry and abundance of Arctic melt ponds can be simply described, which can be exploited in future models of Arctic melt ponds that would improve predictions of the response of sea ice to Arctic warming. 

\end{abstract}

\maketitle

Arctic sea ice plays a major role in Arctic climate \cite{perovich2009loss}, ecology \cite{grebmeier1995biological}, and economy. Sea ice's recent rapid decline is a hallmark of climate change \cite{serreze2007perspectives} that global climate models have systematically underestimated \cite{stroeve2007arctic}. This is believed to be largely due to small-scale processes that cannot be captured accurately by large-scale models \cite{holland1999role}. One such process is the formation of melt ponds on the ice surface during the summer \cite{holland2012improved}. Melt ponds absorb significantly more sunlight than the surrounding ice, making ponded ice melt faster, creating a positive feedback \cite{perovich1996optical,morassutti1996albedo}. The central importance of melt ponds was demonstrated in 2014 by Schroeder et al. \cite{schroder2014september} who showed that the September sea ice minimum extent can be accurately predicted solely based on spring melt pond fraction. Current models of melt ponds include comprehensive representations of many physical processes and are capable of reproducing Arctic-scale spatial distributions of pond coverage \cite{luthje2006modeling,taylor2004model,flocco2007continuum,skyllingstad2009simulation}. However, their complexity and numerous assumptions reduce their ability to provide a fundamental understanding of pond evolution, and call into question their applicability in a changing climate.

Ponds typically evolve through several stages that are controlled by ice permeability \cite{polashenski2012mechanisms,landy2014surface}. Early in the season (typically late spring and early summer), ice is impermeable so that melt ponds can exist above sea level and cover a large portion of the ice. Later in the season, as ice permeability increases, the ponds drain to the ocean so that remaining ponds correspond to regions of sea ice that are below sea level. After drainage, ponds have a typical length-scale of several meters, likely determined by the scale of winter snow dunes \cite{petrich2012snow}, and are often connected by channels that form during drainage. This post-drainage stage is typically the longest part of melt pond evolution. An aerial photograph of drained melt ponds is shown in Figure 1a.

 \begin{figure*}
\centering
\includegraphics[width=0.85\textwidth]{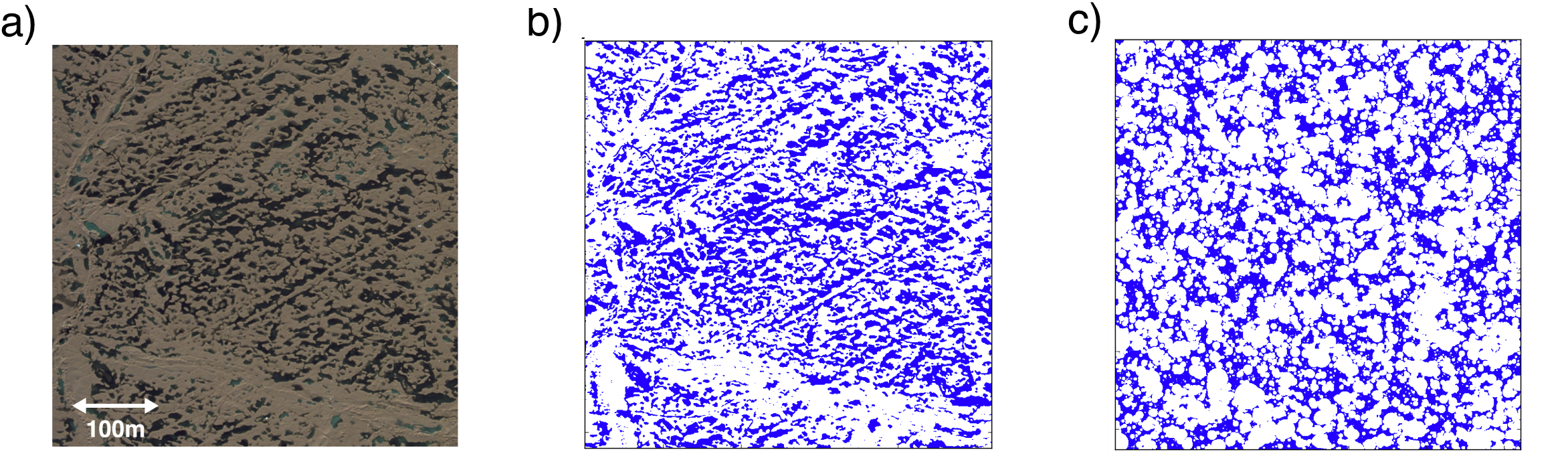}
      \caption{a) A photograph of melt ponds taken on August 7, 1998 during the SHEBA mission. b) A binarized version of the same image. c) A void model with a typical circle radius of $r_0 = 1.8 \text{ m}$, and a coverage fraction of $\rho = 0.31$. }
 \label{fig:1}
 \end{figure*}

Melt pond geometry has been shown to control the strength of lateral melting of ice by pond water \cite{skyllingstad2007numerical}, to impact the pattern of floe breakup \cite{arntsen2015observations}, and to set the landscape of available light for the organisms living beneath the ice \cite{frey2011spatial}. Several critical observations have previously been made about pond geometry. Hohenegger et al. \cite{hohenegger2012transition} showed that the fractal dimension, $D$, of late-summer melt ponds, which characterizes their area-perimeter relationship ($P \propto A^{D/2}$), transitions from $D \approx 1$ for small ponds to $D \approx 2$ for large ponds. The size (area) distribution of melt ponds has also been shown to be a power law \cite{perovich2002aerial}. Although several models reproduce these observations \cite{bowenmodeling,ma2014ising}, a basic understanding of the reason for this behavior is lacking. In this paper we will explain both of these observations using a simple geometric model without invoking any assumptions about the dynamics that govern the melt pond evolution.

Our model is a representation of post-drainage melt ponds. It consists of randomly placing circles of varying size on a plane and allowing them to overlap. The area covered by circles in our model represents ice, while melt ponds are represented by the voids left between the circles (Figure 1c). Similar models are sometimes used to study transport properties in inhomogeneous materials, and are known as ``Swiss cheese'' models \cite{halperin1985differences}. Physically, the circles can be thought of as regions where snow dunes used to be in the winter, and melt ponds fill in the space around them. Circle centers are placed with equal probability throughout the domain. Individual circles have radii, $r$, randomly drawn from an exponential probability distribution $p(r) = \frac{1}{r_0} e^{-r/r_0}$, where $r_0$ is the mean circle radius and defines the physical scale for the model. We chose this probability distribution mainly due to its simple form, but all of our main conclusions are robust to using other distributions (see Supplementary section S4). After choosing $r_0$, the model is fully specified by choosing the fraction of the surface covered by voids, $\rho$. To compare our model with melt pond data, we analyzed hundreds of photographs of sea ice taken during helicopter flights on multiple dates during the SHEBA mission of 1998 and the HOTRAX mission of 2005, and separated them into ice and pond categories using a machine learning algorithm (Figures 1a and b, Supplementary section S1). In order to facilitate comparison with pond images, we implemented the void model on a grid with the same resolution and size as the pond images. 

\begin{figure*}
\centering
\includegraphics[width=0.95\textwidth]{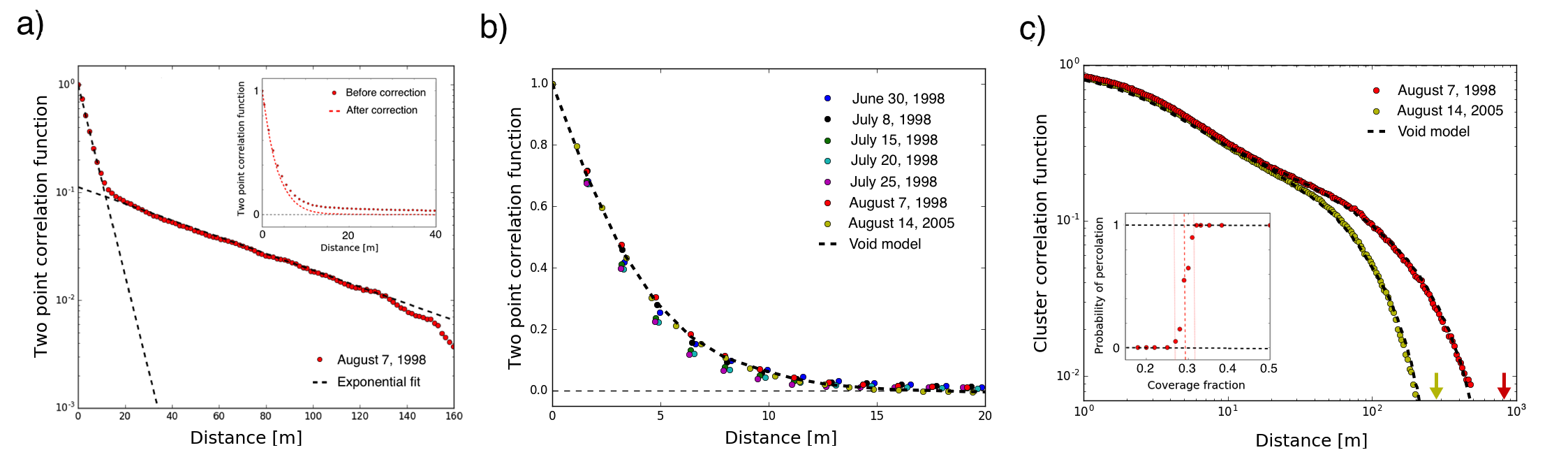}
  \caption{a) An example of the two-point correlation function, $C(l)$, for melt ponds shown on a semi-log plot. Dashed black lines represent fits to a small length scale exponential and a large length scale exponential. The inset shows $C(l)$ before and after a fit to the large length scale exponential has been subtracted.  b) A comparison between the two-point correlation function for ponds from 1998 and 2005 (circles), and the void model (dashed line). Ponds on all dates show a similar scale matched by the void model using $r_0 = 1.8 \text{ m}$. c) A comparison between the cluster correlation function, $g(l)$, for August 7, 1998 (red circles), August 14, 2005 (yellow circles), and the void model using the same $r_0$ as in panel b (black dashed lines). Both model lines use $\rho = 0.31$, and the difference between them is due only to differing simulated image sizes. The image size for 1998 is indicated by a red arrow and the image size for 2005 is indicated by a yellow arrow. The fact that the exponential cutoff is set by the image size indicates that the ponds are roughly at the percolation threshold. The inset shows an independent estimation of the percolation threshold. Red points show the probability of finding a spanning cluster in the void model implemented on a grid the same size and resolution as the SHEBA images. The probability of finding a spanning cluster increases from 0 to 1 between $\rho = 0.28$ and $\rho = 0.31$.}
  \label{fig:2}
 \end{figure*}

We begin the comparison by choosing the model parameters, $r_0$ and $\rho$. To this end, we define two functions - the two-point correlation function, $C(l)$, and a cluster correlation function, $g(l)$, and compare them for pond images and the model. A two-point correlation function measures the probability that two points separated by a distance $l$ are both located on \textit{some} pond, while a cluster correlation function measures the probability that they are both located on \textit{the same} pond. We first estimate $r_0$ using $C(l)$, because we can define it to be largely insensitive to changes in $\rho$ (see below). Once we have calibrated $r_0$ by matching $C(l)$, we can choose $\rho$ using $g(l)$.

For two points, $\textbf{x}$ and $\textbf{y}$, separated by a distance $l$, the two-point correlation function can be defined as: 
\begin{equation}
C(l) = \frac{\langle z(\textbf{x})z(\textbf{y}) \rangle - \rho^2}{\rho(1-\rho)} \text{  ,}
\end{equation}
where $z(\textbf{x}) = 1$ if a point $\textbf{x}$ is located on a pond, and $z(\textbf{x}) = 0$ otherwise, and $\langle...\rangle$ represents averaging over different points and over different images. Subtracting $\rho^2$ and dividing by $\rho(1-\rho)$ constrains $C(l)$ to vary between 1 and 0, and makes it insensitive to changes in $\rho$ (see Supplementary section S2). The two-point correlation function determines a typical length scale of variability in melt pond coverage. 

Plotting $C(l)$ for melt ponds on a semi-log plot reveals that it is approximately a sum of two exponentials (Figure 2a). Therefore, there are two characteristic length scales in melt pond images - a small length scale comparable to the size of individual ponds and a large length scale that is comparable to the size of the image. The large length scale corresponds to variability of pond fraction due to large-scale ice features such as ridges or rafted ice floes. To focus on melt pond features, we have removed the contribution to $C(l)$ from large scale ice features by subtracting a fit to an exponential of $C(l)$ for $l>25 \text{ m}$. We varied this threshold, but found little difference in the results. After subtracting the fit, we normalized the remainder so that $C(0) = 1$ (inset of Figure 2a). We show the resulting functions for all of the available dates and compare them to the void model in Figure 2b. Ponds of all dates show similar $C(l)$ dropping by a factor of $e$ after roughly $3.3 \text{m}$. We found that this is well reproduced by the void model using $r_0 = 1.8 \text{ m}$ (see Supplementary section S2). The fact that the void model reproduces the shape of the two-point correlation function suggests that our assumption of randomly placing the circles is reasonable. 

Next, we determine $\rho$. With this parameter, we wish to capture the pond geometric features such as the pond size distribution and the fractal dimension, rather than simply the pond coverage. For this reason, we do not set $\rho$ equal to the pond coverage fraction of melt pond images, but instead we use the cluster correlation function to determine $\rho$. Essentially, the cluster correlation function, $g(l)$, measures the probability that two points separated by a distance $l$ belong to the same finite pond. However, there are some technical subtleties in how we define $g(l)$, and we give a precise definition in Supplementary section S2.

In the model, in the limit of infinite domain size, there exists a well-defined coverage fraction, $\rho_{\text{c}}$, the ``percolation threshold,'' above which infinite clusters exist, and below which there is a maximum cluster size. The cluster correlation function in the void model sensitively depends on the deviation of the pond fraction from this percolation threshold, $|\rho-\rho_\text{c}|$ (see Supplementary section S2). Below and above the percolation threshold, the cluster correlation function is greater than zero up to a certain distance, after which it exponentially decreases. As the coverage fraction approaches the percolation threshold, this cutoff length grows, and sufficiently close to the threshold, it is set by the image size. The location of the exponential cutoff quantifies the typical size of the largest finite connected pond cluster. We discuss the functional form of $g(l)$ in detail in Supplementary section S6. 

Narrow connections between ponds are often missed by the image processing algorithm so that for many dates $g(l)$ depends on the artificial threshold parameter used in the machine learning algorithm to separate ice from ponds (see Supplementary section S1, for details). The only dates after pond drainage for which $g(l)$ is stable against changes in this threshold are August 7 of 1998 and August 14 of 2005. In Figure 2c, we compare the cluster correlation function for the void model and data on those dates. Remarkably, the pond clusters for both dates appear to be organized very near the percolation threshold, as indicated by the fact that the length scale of exponential cutoff in $g(l)$ is set by the image size. In Figure 2c we use $\rho = 0.31$ to match the pond data, and the difference between $g(l)$ for the ponds from 1998 and ponds from 2005 is solely due to a different image size. In fact, using any $\rho$ from a range $0.28 < \rho < 0.31$ provides an equally good fit to the data, which indicates that within this entire range the size of the largest pond is determined by the image size. To independently confirm that ponds are well-described by the void model near the percolation threshold, we ran the void model, 50 times at multiple values of $\rho$, and found the probability of forming a cluster that spans at least one dimension of the image (inset of Figure 2c). We found that this probability increases from 0 to 1 between $\rho = 0.28$ and $\rho = 0.31$, which closely matches the range of coverage fractions that fit the pond $g(l)$. We note that although we chose $\rho$ to match the cluster structure between the model and the data, the value we found agrees reasonably well with the pond coverage fraction on those dates ($30\% \pm 5\%$ on August 7 of 1998, and around $40\% \pm 5\%$ on August 14 of 2005). We discuss the relationship between the pond coverage fraction and pond geometry in detail in Supplementary section S6. 

It is remarkable that the properties of ponds from 1998 and 2005, which likely developed under very different environmental conditions, are so similar: the correlation functions for both years are well-fit by the void model using the same $r_0$ and $\rho$. This is particularly surprising since sea ice during the 1998 mission had a large proportion of multiyear ice, whereas ice during the 2005 mission was predominantly first-year ice.

% non-spanning
 
 \begin{figure*}
\centering
\includegraphics[width=0.81\textwidth]{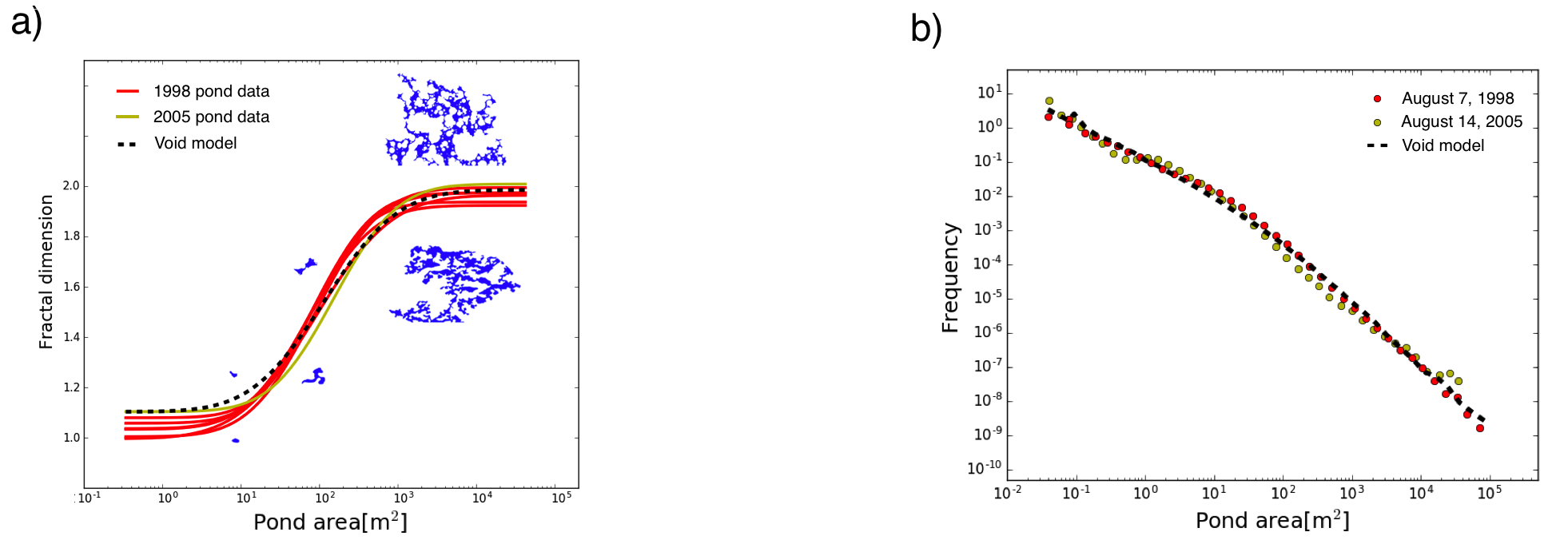}
  \caption{a) A comparison between the fractal dimension of pond boundaries for different dates after pond drainage from 1998 (red curves), 2005 (yellow curve), and the void model with $r_0$ and $\rho$ the same as in Fig 2 (black dashed curve). Examples of ponds (below the curve) and voids (above the curve) of various sizes are also shown. b) Size distribution for ponds on August 7, 1998 (red dots), ponds on August 14, 2005 (yellow dots), and the void model (black dashed line).}
  \label{fig:2}
 \end{figure*}

Having chosen $r_0$ and $\rho$, we can proceed to explain the observations of pond fractal dimension and size distribution. Following Hohenegger et al. \cite{hohenegger2012transition}, we define the fractal dimension of the pond boundary as the exponent that relates the area and the perimeter of the pond, $P \propto A^{D/2}$. The fractal dimension can vary between the fundamental limits of $D = 1$ for regular shapes such as circles to $D = 2$ for space-filling or linear shapes. We find $D$ as a function of $A$ by fitting a curve to the area-perimeter data. We explain the details of this fitting procedure in the Supplementary section S3. 

In Figure 3a we find $D$ as a function of $A$ for pond data on all dates from the summer of 1998 after pond drainage (red curves) and 2005 (yellow curve). Our results are consistent with Hohenegger et al. \cite{hohenegger2012transition}, with the pond fractal dimension transitioning from $D \approx 1$ to $D \approx 2$ at $A_c \approx 100 \text{ m}^2$, and a transition range spanning roughly 2 orders of magnitude. Without any tuning other than choosing $r_0$ and $\rho$ using the correlation functions, the void model is able to match the observed transition in pond fractal dimension nearly perfectly (Figure 3a, black dashed curve).

In the Supplementary section S7, we give an argument that a transition from $D < 2$ to $D \approx 2$ is a generic consequence of individual objects connecting and, therefore, cannot be used as strong support for any particular physical model of melt ponds. On the other hand, matching the fractal transition scale and the transition range are non-trivial, and cannot be reproduced by an arbitrary model of randomly connecting objects (see Supplementary section S9). At small sizes, the void model predicts a dimension slightly larger than 1, likely corresponding to the fact that small voids are not necessarily simple smooth shapes. It is possible that small-scale physical processes in real ponds, such as erosion of pond walls, are responsible for smoothing small ponds into more circular shapes with $D \approx 1$.

Finally, we compare the pond size distribution with the void model in Figure 3b. Again as a result of sensitivity to the threshold parameter in the machine learning algorithm, we only use pond data for August 7th of 1998 and August 14th of 2005. At scales larger than roughly 10 $m^2$ the pond size distribution follows an approximate power law, in agreement with previous findings. The power law behavior is particularly clear for ponds from 1998, and the power law exponent (approximately 1.8) is slightly larger than previously found \cite{perovich2002aerial}. Using the same $r_0$ and $\rho$ as before, the void model reproduces the pond size distribution over the entire range of observations, more than 6 orders of magnitude. This matching is highly robust: the void model matches the pond size distribution even at the smallest scales regardless of details such as the circle radius distribution or the shape of the objects placed randomly (see Supplementary section S4). 

We have shown that a simple model of voids surrounding overlapping circles captures key geometric patterns of Arctic melt ponds with high fidelity and robustness, with only two parameters that can be chosen naturally by comparing the model and the data. Our model is purely geometric, and can therefore be used as a benchmark against which to test any physical model. This work shows that much of melt pond geometry can be understood simply by assuming that melt ponds are placed randomly and have a typical size. Even though many models will reproduce the same universal features, our model is special in that it captures quantitative details of melt pond geometry beyond what an arbitrary model of connecting objects is capable of doing. Our work raises two critical questions about melt pond physics that must be answered. First, why does the pond scale appear to be so robust for ponds evolving under differing environmental conditions, and, second, why do ponds seem to be organized near the percolation threshold? The answer to the second question may be particularly interesting, as it may point to self-organized critical behavior in melt ponds, and may suggest that the pond coverage fraction is more constrained than previously thought. Answering these questions may yield deeper insight into melt pond physics and allow for a better representation of this important process in global climate models.%\nocite{SupplementaryVoidModel,mandelbrot1967long,aharony2003introduction,essam1980percolation,goldenfeld1992lectures,delfino2010universal,bowenmodeling,ma2014ising,kerstein1983equivalence,sicilia2007domain}

\textbf{Acknowledgements:} We thank Don Perovich for providing the image data. We thank Djordje Spasojevi\'{c} and Alberto Petri for discussions about correlation functions and the percolation threshold. We thank Douglas MacAyeal for reading the paper and giving comments. We thank three anonymous reviewers for constructive comments. Predrag Popovi\'{c} was supported by a NASA Earth and Space Science Fellowship. B. B. Cael was supported by National Science Foundation Graduate Research Fellowship Program, grant number 2388357. This work was partially supported by the National Science Foundation under NSF award number 1623064.

\bibliographystyle{apsrev4-1}
\bibliography{mybibliog}{}

\end{document}

\renewcommand{\thetable}{S\arabic{table}}
\renewcommand{\thefigure}{S\arabic{figure}}
\renewcommand{\thesection}{S\arabic{section}}
\renewcommand{\theequation}{S\arabic{equation}}

\maketitle

\section{Image analysis}
We used airborne photographs taken during the SHEBA mission of 1998 and the HOTRAX mission of 2005 (Figure 1a of the main text). During SHEBA, sea ice was regularly photographed from a helicopter, and the SHEBA photographs are available on eight dates spanning the entire melt season of 1998. Six of those eight dates were after pond drainage. Helicopter photographs from the HOTRAX mission that have unfrozen ponds on unbroken ice floes are only available for August 14th of 2005. SHEBA images have dimensions of 819 m by 1228 m, with a resolution of roughly 0.2 m per pixel. HOTRAX images have dimensions of 427 m by 284 m, with a resolution of 0.14 m per pixel and are higher quality than SHEBA images. For each available date, we only used images that have few cracks in the ice and little crushed ice that might be mistaken for melt ponds by the image classifying algorithm. We also manually removed the regions of open ocean before running the algorithm. Finally, we separated ice from ponds using an open access machine learning software ``ilastik'' (http://ilastik.org/). For most dates we analyzed more than $10^5$ individual ponds. 

We trained the machine learning algorithm using information from images about color, intensity, color gradient, and texture. The output of the algorithm is a probability matrix characterizing the likelihood that each pixel is ice or pond. To identify ponds, we chose a threshold probability, $p_t$ (usually 0.5), and classified each pixel as a melt pond if the algorithm found it to have a higher probability than $p_t$. To make sure melt pond features we wish to describe are robust, we varied the threshold probability. We found that some characteristics of binarized images depend on $p_t$. For example, pond coverage fraction varies by as much as $10 \%$ between $p_t = 0.1$ and $p_t = 0.9$. For this reason, in the main text and the remainder of the Supplementary Material, when referring to the mean pond coverage fraction on particular dates, we also give a range of coverage fractions that can be obtained by changing $p_t$. Furthermore, we found that for many dates the cluster correlation function, $g(l)$, and the pond size distribution are sensitive to this parameter. Therefore, we only considered $g(l)$ and the pond size distribution on dates for which ponds could be clearly distinguished so that these statistics were insensitive to changes in $p_t$. This was true only for June 22 and August 7 of 1998 and August 14 of 2005. As June 22 is before complete pond drainage, in the main text we report $g(l)$ and the pond size distribution only for August 7 of 1998 and August 14 of 2005. This sensitivity of $g(l)$ and the size distribution to $p_t$ is likely due to the fact that these statistics rely on accurately identifying the narrow connections between the ponds. Images from 2005 were high enough quality, and August 7 of 1998 had ponds that were dark enough to be easily distinguished from ice. However, we acknowledge the possibility that the perceived similarity in the cluster correlation function for the two dates may be due to this early selection bias.

% Because we were not able to test other dates, we cannot say when ponds become organized close to the percolation threshold.
%However, this sensitivity could also be a sign that the geometric characteristics of melt ponds on the other days differ in significant ways with possible implications for why the two dates are so close to the percolation threshold. 

\begin{figure}
  \centering
   \includegraphics[width=1\textwidth]{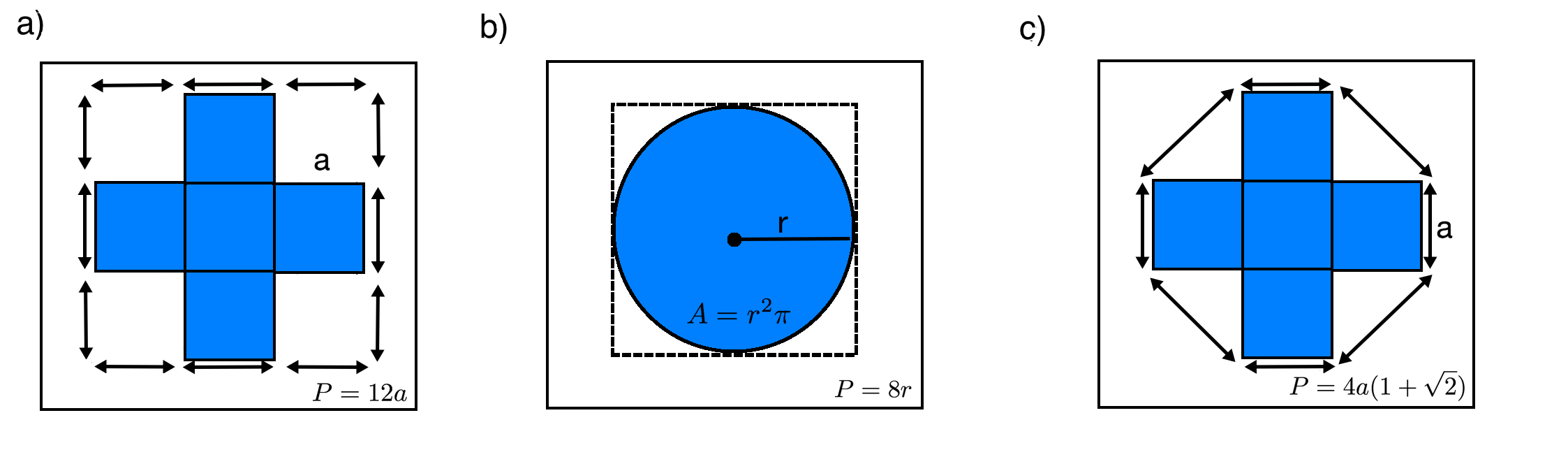}
   \linespread{1.}\selectfont{}
\caption{A diagram explaining how we find the perimeter. a) Blue squares represent individual pond pixels. The perimeter is estimated as the sum of all boundary pixel edges. b) Finding the perimeter of a circle as in panel a), we inaccurately estimate it to be the perimeter of a square surrounding the circle. Estimating the perimeter in this way is equally inaccurate regardless of how fine the image resolution is. c) The problem is partially corrected if we take into account the relative positions of the nearest neighbor boundary pixels. If two nearest neighbor pixels are located diagonally, we add the distance between them to the total perimeter.}
  \label{fig:1}
 \end{figure}

The size of melt ponds can be accurately estimated by summing all the areas of individual pixels within a pond. Estimating the pond perimeter is slightly more challenging (Figure \ref{fig:1}). Summing the lengths of all pixel edges on the pond boundary gives an inaccurate estimate of the perimeter, because pixels are located on a grid, and cannot take into account the curvature of the boundary. For example, if we try to estimate the perimeter of a circle by summing the lengths of all pixel edges on its boundary, we will get the perimeter of a square enclosing that circle (Figure \ref{fig:1}b). We partially correct for this by considering the nearest neighbor pixels on the boundary: if the nearest neighbors are positioned diagonally, we add a distance between them to the total perimeter (Figure \ref{fig:1}c). Even with this correction, there is still a small systematic error in the estimate. This, however, did not affect our estimates of the pond statistics: different methods used for finding the perimeter simply introduced a constant bias in the perimeter of the ponds, and therefore did not change our estimates of the fractal dimension. Some objects in nature (a notable example is the coast of Britain \cite{mandelbrot1967long}) suffer from a fundamental difficulty in determining the perimeter, because the length of the perimeter depends on the length of the measuring stick. In our case this is not a problem, because small ponds are regular shapes ($D_1 \approx 1$) and we can resolve them easily in our images. 

%Since small ponds are regular shapes ($D_1 \approx 1$), and because the size of a pixel is much smaller than the typical size of individual ponds, pond perimeter, unlike the coast of Britain \cite{mandelbrot1967long}, does not depend significantly on image resolution.

\section{Correlation functions}

\begin{figure}
  \centering
   \includegraphics[width=1\textwidth]{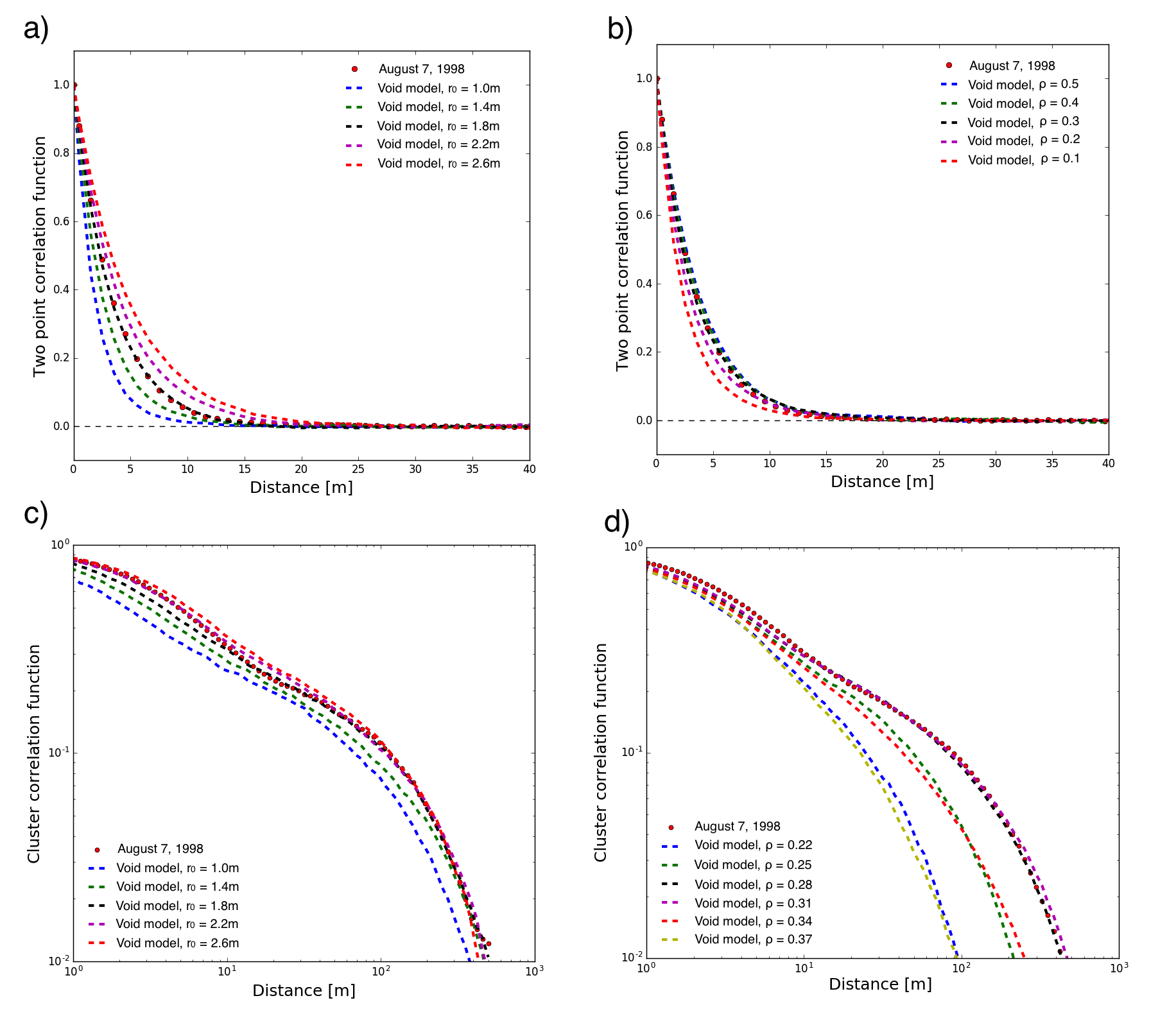}
   \linespread{1.}\selectfont{}
  \caption{Dependence of correlation functions on model parameters $r_0$ and $\rho$. In each plot, red dots represent data for August 7th,1998. a) Two-point correlation function for different values of $r_0$ at $\rho = 0.3$. b) Two-point correlation function for different values of $\rho$ at $r_0 = 1.8 \text{m}$. c) Cluster correlation function for different values of $r_0$ at $\rho = 0.3$. d) Cluster correlation function for different values of $\rho$ at $r_0 = 1.8 \text{m}$. }
  \label{fig:3}
 \end{figure}

In this section, we will first give a precise definition of the cluster correlation function, $g(l)$, and then we will explore how the two correlation functions, $C(l)$ and $g(l)$, depend on the model parameters $r_0$ and $\rho$.

We define the cluster correlation function, $g(l)$, as the probability that two points separated by a distance $l$ belong to the same non-spanning pond given that one of the points is already located on a non-spanning pond. For both model and data, ``spanning ponds'' are those ponds that span at least one dimension of the image. In order to obtain a good fit to the data, it is necessary to exclude spanning clusters from the computation of $g(l)$. This is reasonable since in the pond images, ponds are constrained by large scale features, such as floe edges or ridges, which make the void model inapplicable above a certain pond size (see also Supplementary section S5). In the void model, above the percolation threshold, there is typically one spanning pond, while in the pond images spanning ponds typically do not exist.

We found the parameters $r_0$ and $\rho$ approximately by running the model at multiple values of of these parameters, and among these runs choosing the one for which the two-point correlation function, $C(l)$, and the cluster correlation function, $g(l)$, best agree with the correlation functions of ponds. Because $r_0$ is the only length scale in the model (apart from the image size), all of the characteristic lengths must scale with $r_0$. In Figs. \ref{fig:3}a and c, we show how $C(l)$ and $g(l)$ depend on $r_0$. We can see that the decay rate of $C(l)$ for the model is proportional to $r_0$, and the model and data agree well for $r_0 = 1.8 \text{m}$. At small $l$, $g(l)$ is also scaled by $r_0$, but at large $l$, it is insensitive to changes in $r_0$ because the cutoff length is set by the image size. In Fig. \ref{fig:3}b and d, we show how $C(l)$ and $g(l)$ vary with $\rho$ at fixed $r_0$. The two-point correlation function is largely insensitive to changes in $\rho$: $C(l)$ decreases slightly with decreasing $\rho$, but this effect only becomes noticeable at low coverage fraction, $\rho \approx 0.1$, beyond the range of coverage fractions we are considering in this paper. On the other hand, $g(l)$ depends sensitively on $\rho$. The cutoff length of the cluster correlation function reaches its maximum close to the percolation threshold ($\rho \approx 0.3$), and decreases sharply when $\rho$ deviates from this threshold. Because we excluded the spanning clusters from calculations of $g(l)$, the cluster correlation function has the same shape both above and below the percolation threshold.  

%Equation (1) of the main text defining the two-point correlation function, $C(l)$, may be rewritten as 
%\begin{equation}
%C(l) = \frac{p(z(\textbf{y}) = 1 | z(\textbf{x}) = 1) - \rho}{1-\rho} \text{ , }
%\end{equation}
%where $p(z(\textbf{y}) = 1 | z(\textbf{x}) = 1)$ is the probability that a point $\textbf{y}$ is on a pond given that $\textbf{x}$ is on a pond, and $\textbf{x}$ and $\textbf{y}$ are separated by a distance $l$. To estimate $C(l)$ for the model and images, we implement this formula rather than Eq. (1) of the main text. We estimate $p(z(\textbf{y}) = 1 | z(\textbf{x}) = 1)$ by randomly placing a certain number of points on ponds and finding a fraction of points a distance $l$ away that also belong to ponds. When estimating this fraction, we do not take into account ocean points.

%The definition of the two-point correlation function, Eq. (1) of the main text, may be equivalently written as:
%\begin{equation}
%C(l) = \frac{p(z(\textbf{y}) = 1 | z(\textbf{x}) = 1) - \rho}{1-\rho} \text{ , }
%\end{equation}
%where $p(z(\textbf{y}) = 1 | z(\textbf{x}) = 1)$ is the probability that a point $\textbf{y}$ is on a pond given that $\textbf{x}$ is on a pond. The two-point correlation function, $C(l)$, is related to the probability of two points separated by a distance $l$ belonging to some pond.

\section{Estimating the fractal dimension}

\begin{figure}
  \centering
   \includegraphics[width=1\textwidth]{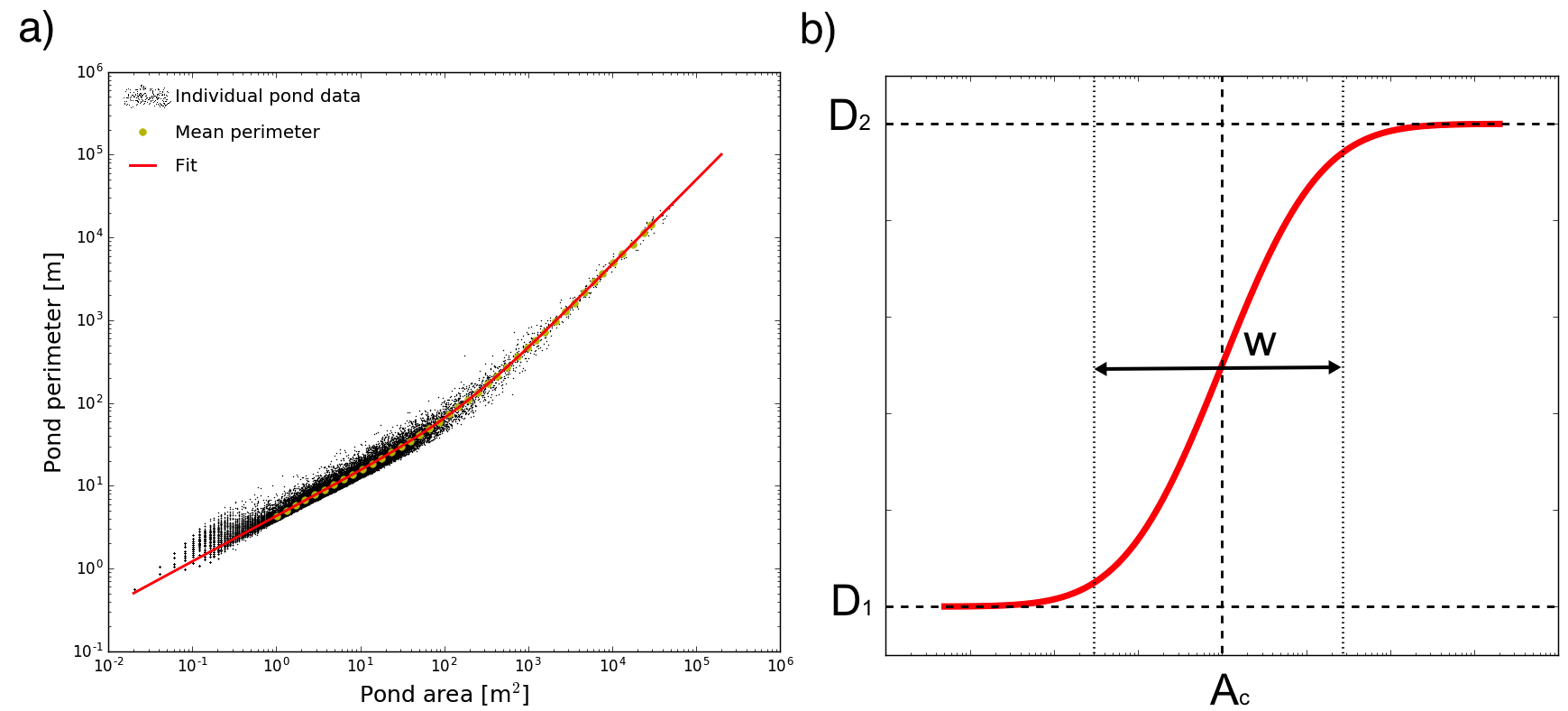}
   \linespread{1.}\selectfont{}
  \caption{An explanation for the fitting procedure to determine the fractal dimension curve. a) The black dots represent area and perimeter of individual ponds. The yellow dots represent a moving average of the perimeter. The red line is a fit of the mean area-perimeter data to a function $f(\log(A))$ defined in Eq. (\ref{fx1}). b) Fractal dimension, $D$, as a function of size, determined as a derivative of the red line in panel a with respect to $\log A$. Fitting parameters defined in Eq. (2) of the main text are also shown. }
  \label{fig:4}
 \end{figure}

The fractal dimension may be obtained from the derivative of the pond perimeter with respect to pond area in log-log space, $D = 2 \frac{d \log{P}}{d \log{A}}$. In order to estimate the fractal dimension, we first find the moving average of the perimeter of all the ponds that fall into a certain log-area bin, $\langle P \rangle$, as a function of $\log{A}$. A log-area bin of width $\Delta$ centered on $A$ is defined as a range from $A / \Delta$ to $A \Delta$. Log-binning defines a set of points $(\log{A},\log{ \langle P \rangle})$. Direct differencing of these data gives noisy results. Instead, we first fit a function to the $(\log{A},\log{\langle P \rangle})$ points, and then take its derivative. Anticipating that the fractal dimension will change from a low to a high value, we choose to represent it with an error function of log-area
\begin{equation}
D(A) = \frac{D_2 - D_1}{2} \text{ erf }\log{(A/A_c)^{1/w}} + \frac{D_2 + D_1}{2} \text{  ,}
\label{DA}
\end{equation}
where $D_1$, $D_2$, $\log{A_c}$, and $w$ are fitting parameters and represent the lower fractal dimension, the upper fractal dimension, the center of the fractal transition in log-area, and the width of the fractal transition. Assuming this form of $D(A)$, the fitting function is given by the integral of Eq. \ref{DA} (Figure \ref{fig:4}a):
\begin{equation} \label{fx1}
f(x) = \frac{D_2 - D_1}{4} \big( (x - \log{A_c}) \text{ erf } \frac{x - \log{A_c}}{w} + \frac{w}{\sqrt{\pi}}e^{-\frac{(x - \log{A_c})^2}{w^2}} + \frac{D_2 + D_1}{D_2 - D_1}x  \big) + C  \text{  ,}
\end{equation}
where $C$ is a constant of integration, which we regard as another fitting parameter. The fractal dimension is then found as
\begin{equation} 
D = 2 \frac {d f(\log(A))}{d \log{A}} \text{ ,}
\end{equation} 
and is given by Eq. \ref{DA} (Figure \ref{fig:4}b). We found that this method of estimating $D$ is in good agreement with other similar methods, such as fitting a function to directly differenced data or smoothing out the directly differenced data. Log-binning the data before any additional processing was important for two reasons: 1) it smooths the data, and 2) it assigns equal weights to large and small ponds during fitting. This latter property is especially important since there are several orders of magnitude more small ponds than large ponds, and a fit to data that was not log-binned would be determined nearly entirely by small ponds. 

%The fractal dimension is then found as
%\begin{equation} 
%D = 2 \frac {d f(\log(A))}{d \log{A}} \text{ ,}
%\end{equation}
%and is given by Eq. (2) of the main text (Figure S2b).

\section{Robustness of the void model}

The void model reproduced the pond data highly robustly, regardless of details such as the distribution of circle radii, $p(r)$, or the exact shape of the objects placed. In addition to an exponential distribution of circle radii, we tested the model using other distributions, such as Gaussian, Rayleigh, and Gamma distributions, and found that the correlation functions, fractal dimension, and size distribution are insensitive to these details. Even in the limit of no variation in the circle radius, all of the characteristics can be reasonably well reproduced, although the agreement with the data is affected somewhat. Real melt ponds are often strongly anisotropic (compare Figures 1b and c of the main text). We tested the effect of anisotropy in our model by placing randomly sized ellipses with a fixed ratio of semi-major to semi-minor axis instead of circles. The ellipses had a preferred orientation and we changed the degree to which they align with this preferred orientation. None of these changes altered the main conclusions of our model: matching the correlation functions led to matching fractal dimension curves and size distributions for the broad range of ellipse parameters used in the simulations. Some quantities were slightly affected. For example anisotropy decreased the value of the percolation threshold by several percent.

% Additionally, we tested a model of voids around randomly sized and placed ellipses with a fixed ratio of semi-major to semi-minor axis.
%  if there are winds that determine the orientation of the snow dunes

In addition to the void model, we also explored its negative: a model where ponds are represented by overlapping circles. The circle model can also reproduce most of the observations, but the results are less robust. For example, matching the two correlation functions does not reproduce the center of the fractal transition. It is possible to remedy this by placing ellipses instead of circles, but in this case the range of the transition is affected by the ratio of the semi-major to semi-minor axis. The circle model also suffers other drawbacks compared to the void model: it reproduces the pond size distribution over only 4, rather than more than 6, orders of magnitude, it does not match the cluster correlation function as well, and it matches the observations at a $\rho$ significantly higher than the actual pond coverage (actual pond coverage fractions for August 7th of 1998 and August 14th of 2005 were $0.3 \pm 0.05$ and $0.4 \pm 0.05$, close to the void model $\rho \approx 0.3$, and significantly less than the circle model $\rho \approx 0.7$ ). For these reasons, we believe the void model is a better description of melt ponds than the circle model. 

The exponent of 1.8 we found for the pond size distribution is slightly less than 2.05, which is predicted for the universality class of percolation models. This is likely due to finite size effects, as the image size is less than two orders of magnitude larger than the length scale determined by the two point correlation function. To support this hypothesis, we ran a ``site percolation'' model for different lattice sizes. In this model, each grid point on a lattice is occupied with a certain probability, and two occupied nearest neighbor sites are considered connected. We found that the cluster size distribution of site percolation on a 100x100 lattice decays with an exponent close to 1.8, while the exponent approaches 2.05 for large lattices.

\section{Limitations of the void model}

\begin{figure}
  \centering
   \includegraphics[width=1\textwidth]{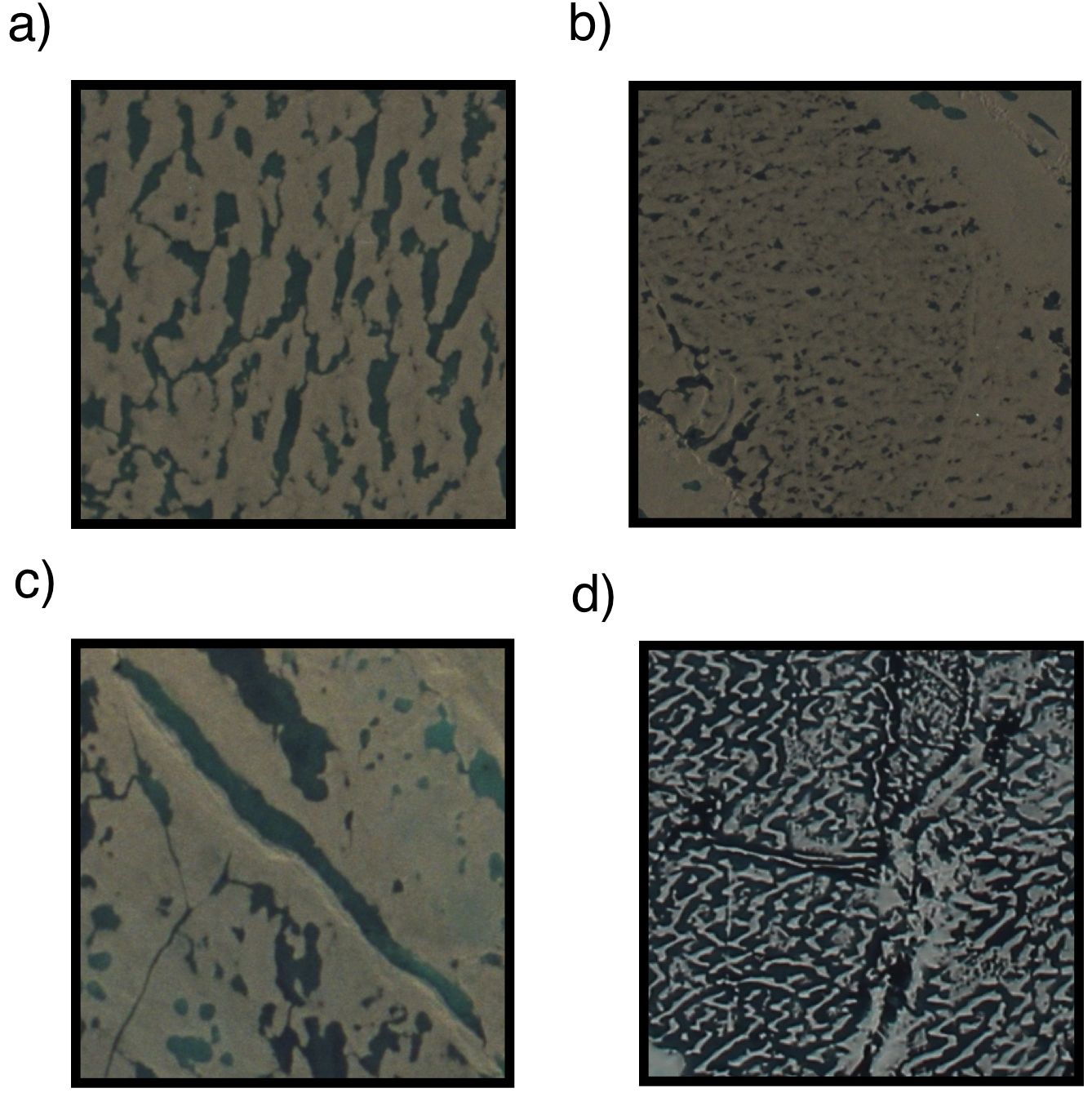}
   \linespread{1.}\selectfont{}
  \caption{Examples of melt ponds that appear to violate the assumption of random placement. a) ``Banded'' melt ponds with clear spacings between melt pond bands. b) A low pond coverage region of the ice with small melt ponds that seem to be located non-randomly. c) A long melt pond located along a ridge. d) ``Tiger stripe'' melt ponds with clear ordering. }
  \label{fig:S4}
 \end{figure}

When calibrating the circle scale using the two-point correlation function, we had to remove a long length-scale exponential from the correlation function. This indicates a limitation of our model: it is unable to represent pond variability on an arbitrarily large scale, because in real ice there exist large features, such as ridges, cracks, or floe edges, that are not represented in the model. One result of this limitation is that the void model predicts infinite ponds, which are unrealistic, and which we had to remove from the analysis in order to obtain a good match between the model and the data (see the definition of the cluster correlation function in section S2). The limit of applicability of our model can be estimated from the long length exponential of the two-point correlation function to be several hundred meters. 

One of the key assumptions of our model is the random placement of circles on a plane; however, real melt ponds sometimes violate this assumption. Examples of ponds that are not randomly placed are shown in Figure \ref{fig:S4}. The fact that our model is able to reproduce pond statistics suggests that these types of ponds were not very prevalent in our data. However, it may happen that under different conditions, non-random ponds might become significant. 

%``Snow dunes'' in our model are represented as circles, which leads to modeled ponds being isotropic. Real melt ponds, however, are often strongly anisotropic if there are winds that determine the orientation of the snow dunes (compare Figures 1b and c of the main text). We tested the effect of anisotropy in our model by placing ellipses with a preferred orientation instead of circles. As long as ellipses were placed randomly, anisotropy did not have an effect on the shapes of the two correlation functions, the fractal dimension, and the size distribution. It did, however, have some effect on the coverage fraction needed to achieve the percolation threshold, $\rho_c$, and on the precise location of the fractal transition.

\section{Relationship between pond geometry and coverage fraction}

\begin{figure}
  \centering
   \includegraphics[width=1\textwidth]{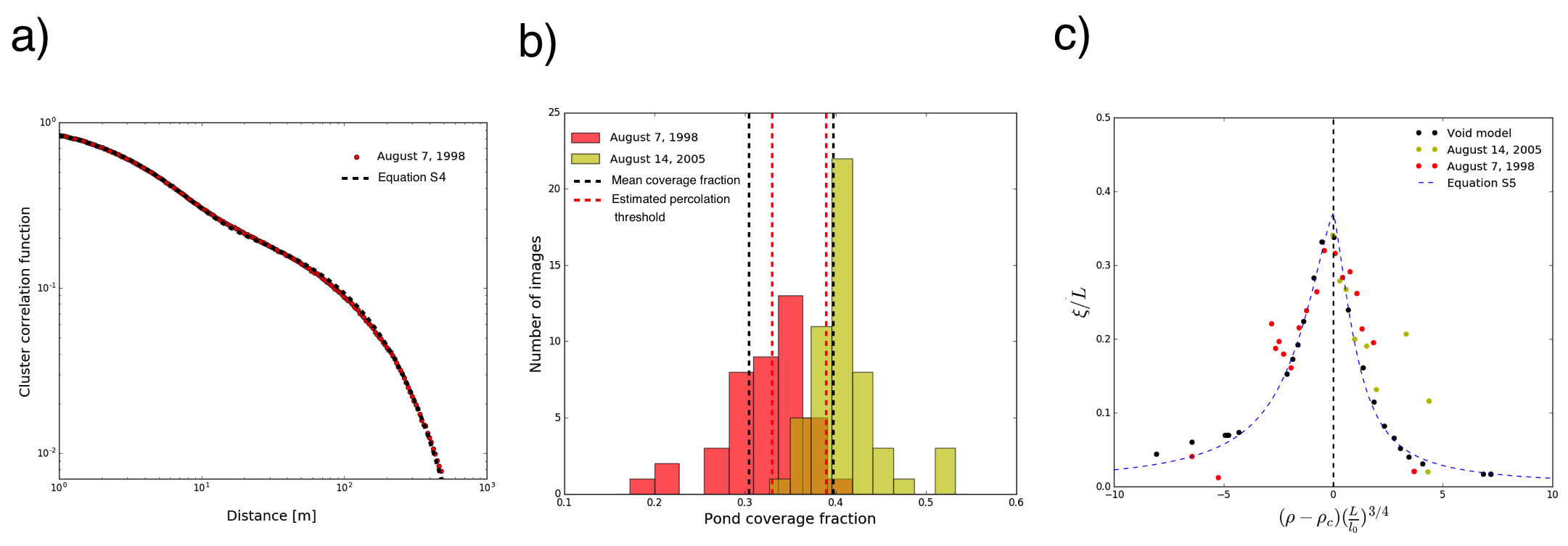}
   \linespread{1.}\selectfont{}
  \caption{ a) A fit of Eq. \ref{gl} (black dashed line) to the cluster correlation function of ponds on August 7, 1998 (red dots). We set $\rho$ equal to the mean pond coverage on August 7, and we treat $l_0$ and $\xi$ as fitting parameters. b) Number of images that fall into each bin of pond coverage fraction for August 7, 1998 (red bars) and  August 14, 1998 (yellow bars). Vertical black dashed lines represent mean coverage fraction on the two dates, while the vertical red dashed lines represent the estimated coverage fraction of the percolation threshold for each date. All of the pond coverage fractions were estimated using the machine learning threshold $p_t = 0.5$. c) Estimated correlation length scaled by the image size, $\xi/L$, as a function of the appropriately scaled distance from the percolation threshold, $(\rho-\rho_c)(L/l_0)^{3/4}$, for the void model (black dots), August 7, 1998 (red dots) and  August 14, 2005 (yellow dots). Values for the percolation threshold used were 0.3 for the void model, 0.33 for 1998 ponds, and 0.39 for 2005 ponds. Also shown is Eq. \ref{xi} (blue dashed line), consistent with theoretical considerations. }
  \label{fig:S5}
 \end{figure}

The parameter $\rho$ in the void model controls both the connectedness of the voids and the void coverage fraction. A priori, there is no reason to believe that such a link between coverage fraction and geometry exists in real melt ponds. For example, ponds may be connected by narrow channels, thereby increasing the typical pond size while leaving the coverage fraction virtually unchanged. On the other hand, pond growth by lateral melting would likely increase the coverage fraction without changing the connections between ponds much. In the main text, we chose $\rho$ in the void model such that it reproduces the geometry of melt ponds, and did not consider the pond coverage fraction and whether it is related to geometry. For this reason, in this section we will show that this relationship also exists in real melt ponds. To avoid confusion, in this section we will call $\rho^{m}$ the coverage fraction in the model, and $\rho^{p}$ the pond coverage fraction.

The key to understanding pond connectedness is the cluster correlation function, $g(l)$. We can guess the functional form of $g(l)$ for the void model solely from considering its asymptotics. Properties of $g(l)$ to note are:
\begin{enumerate}
\item At small separation, $l$, two points located on a pond will most likely belong to the same pond. So, $g(l)$ should be the same as the two-point correlation function up to normalization by the coverage fraction. Therefore, $g(l) \approx (e^{-l/l_0}(1-\rho) + \rho)$ for $l \ll l_0$, where $e^{-l/l_0}$ is the two-point correlation function for randomly placed objects.
\item At separations on the order of or larger than the correlation length, $\xi$, $g(l)$ should decay to zero exponentially. So, $g(l) \approx e^{-l/\xi}$ for $l >\xi$. 
\item At intermediate separations, $l_0 \ll l \ll \xi$, percolation theory predicts a power law decay with a universal exponent, $g(l) \propto l^{-5/24}$ \cite{aharony2003introduction,essam1980percolation}.
\end{enumerate}
A function consistent with these asymptotics is:
\begin{equation}
g(l) =  (e^{-l/l_0}(1-\rho) + \rho) (\frac{l_0+l}{l_0})^{-5/24}e^{-l/\xi} \text{  . } 
\label{gl}
\end{equation}
In Figure \ref{fig:S5}a, we show that this equation and $g(l)$ for ponds on August 7th, 1998 agree nearly perfectly. Equally good fits can be obtained for 2005 ponds and for the void model. 

The link between coverage fraction and geometry in the void model is reflected in the fact that the correlation length, $\xi$, that measures the size of the largest voids, is a function of the coverage fraction, $\rho^m$. Near the percolation threshold, $\rho_c$, percolation theory predicts this dependence to be of the form $\xi = \xi_{\infty}f(\frac{L}{\xi_{\infty}})$ \cite{goldenfeld1992lectures,aharony2003introduction,essam1980percolation}, where $L$ is the image size, $\xi_{\infty}$ is the correlation length on an infinite image, and $f(x)$ is a universal function that scales as $f(x) \propto x$ for $x \rightarrow 0$, and $f(x) \rightarrow 1$ for $x \rightarrow \infty$. The correlation length on an infinite image, $\xi_{\infty}$, is given by $\xi_{\infty} = A_{\pm} l_0 |\rho^m - \rho_c|^{-4/3}$ \cite{aharony2003introduction,essam1980percolation}, where $\pm$ stands for $\rho^m > \rho_c$ and $\rho^m < \rho_c$, and $A_{\pm}$ are non-dimensional numbers with $A_{-}/A_{+} = 2$ \cite{delfino2010universal}. To obtain the correct units, the correlation length must be proportional to the fundamental length scale, $l_0$, that can be estimated from a fit to Eq. \ref{gl}. A non-dimensional correlation length, $\hat{\xi} \equiv \xi/L$, consistent with these asymptotics is
\begin{equation}
\hat{\xi} = \hat{\xi}_{\infty}\Big( 1 - e^{- \frac{B}{\hat{\xi}_{\infty}}}   \Big)  \text{ ,}
\label{xi} 
\end{equation}
where $B$ is a non-dimensional number, and $\hat{\xi}_{\infty} = \xi_{\infty}/L$. Here, we wish to test whether there is such a relationship between pond coverage, $\rho^p$, and the correlation length in real melt pond images.

In Figure \ref{fig:S5}b, we show the distribution of pond coverage fraction for images taken on August 7th, 1998 and August 14th, 2005, estimated using the threshold $p_t = 0.5$, used by the machine learning algorithm to identify the melt ponds. We can use the fact that there is substantial spread around the mean pond coverage to test the relationship between $\xi$ and $\rho^p$. We split the entire range of pond fractions into bins and find $g(l)$ only for images with $\rho^p$ that falls into a certain bin. We then use Eq. \ref{gl} to fit $g(l)$ for each of the coverage bins. When fitting, we set the parameter $\rho = \rho^p$, and treat $l_0$ and $\xi$ as fitting parameters. We also perform the same procedure to the void model with different $\rho^m$. 

In Figure \ref{fig:S5}c, we compare the model, the data, and Eq. \ref{xi}. The void model conforms to Eq. \ref{xi} except sufficiently far from the percolation threshold where the theoretical prediction for $\xi_{\infty}$ is no longer valid. Melt ponds on both August 7th, 1998, and August 14th, 2005 also seem to follow the same trend. These data indicate that the pond coverage fraction controls the pond geometry in a similar way as in the void model. However, we cannot simply relate $\rho^m$ in the void model to the pond coverage fraction, because the values of the percolation threshold differ between the two dates and the model. To obtain a match in Figure \ref{fig:S5}c, we used $\rho_c = 0.3$ for the void model, $\rho_c = 0.33$ for 1998 ponds, and $\rho_c = 0.39$ for 2005 ponds. In Figure \ref{fig:S5}b we indicated these values and compared them to the mean pond coverage fraction. For both dates, the percolation threshold used in Figure \ref{fig:S5}c is very close to the mean coverage fraction, indicating again that the ponds are organized near the percolation threshold. 

Mean pond coverage fraction, and its effect on sea ice albedo, is often the main quantity of interest in the large scale models. Here we showed that the mean $\rho^p$ is very close to the percolation threshold, meaning that $\rho^p$ may be more constrained than previously thought. Understanding what physically sets the percolation threshold may be crucial to understanding the evolution of melt ponds and representing them in large scale models. 

%In the void model, $\rho_c$ is a non-universal quantity, meaning that it depends on the model details. For example placing ellipses instead of circles, changes the value of the percolation threshold.  

%Percolation theory gives a prediction for the correlation length as a function of $\rho$ close to the percolation threshold, $\rho_c$. When the size of the image is infinite, the theoretical prediction is:
%\begin{equation}
%\xi = A_{\pm} |\rho - \rho_c|^{-4/3} \text{  , }
%\label{xi}
%\end{equation}
%where $A_{+}$ and $A_{-}$ are constant amplitudes for $\rho > \rho_c$ and $\rho < \rho_c$, and $-4/3$ is a universal critical exponent. Constants $A_{\pm}$ are non-universal (they depend on the details), but their ratio is universal, $A_{-} = 2A_{+}$.  

\section{A fractal transition is a general consequence of connecting objects}

\begin{figure}
  \centering
   \includegraphics[width=0.8\textwidth]{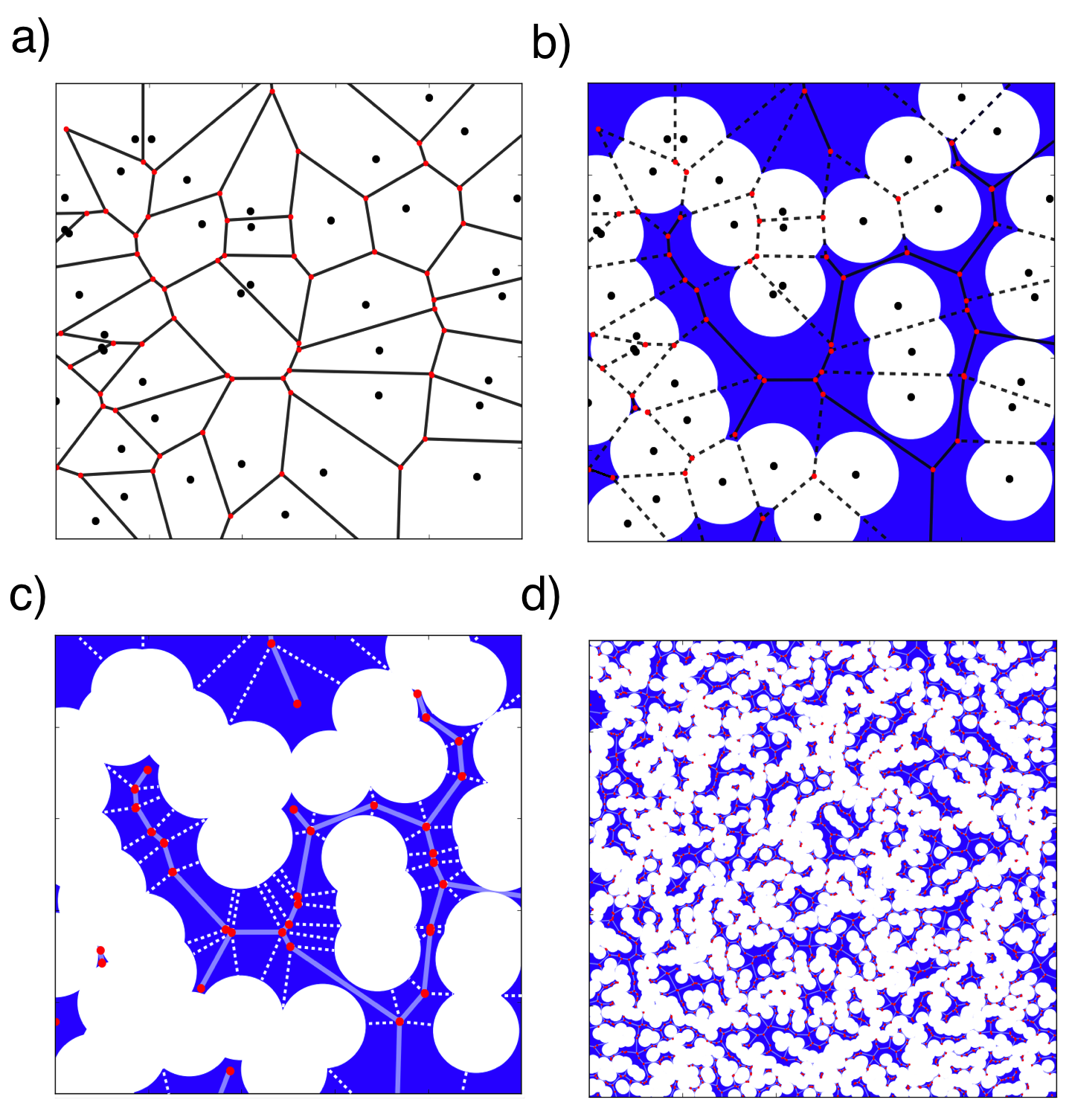}
   \linespread{1.}\selectfont{}
  \caption{a) Circle centers are placed randomly on a plane (black dots). We segment the plane into regions, each region being a set of points closest to a circle center. These regions are polygons and define a natural ``grid'' for the void model. Sides of the polygons are bonds of the grid (black lines), and corners of the polygons are nodes of the grid (red dots). b) Bonds are removed if they pass through a circle. Removed bonds are shown as black dashed lines. The remaining bonds (solid black lines) all lie within voids and connected bonds correspond to connected voids. c) Each void can be partitioned into a sum of contributions from bonds. We can do this in the following way. Every node (red dots) is associated with three circle centers and, correspondingly, three bonds (solid pale blue lines). If all three of these bonds belong to a void, we draw three lines from a node towards its corresponding circle centers. If two bonds belong to a void, but one intersects a circle, we draw two lines from a node - one along the missing bond and one towards a circle center not associated with the missing bond. Finally, if a node only has one bond that belongs to a void, we draw no lines. Lines drawn in this way (white dashed lines) segment a void in a unique way, with each segment associated with only one bond. Contributions to area and perimeter segments vary significantly, but have a typical scale. d) For large enough voids, variability in area and perimeter of segments associated with each bond tends to average out, making both the total area and perimeter proportional to the number of bonds in a void.}
  \label{fig:6}
 \end{figure}

The void model is not the only model able to produce a transition in fractal dimension from $D < 2$ to $D = 2$. As described above, its negative, a model of overlapping circles, produces a transition from $D = 1$ to $D = 2$. Many other models, such as a model of random topography \cite{bowenmodeling} or the Ising model \cite{ma2014ising}, also reproduce the same feature. Here we give a qualitative argument for why this is a general feature of connecting objects. 

To understand why the transition happens in the void model, we will first show that voids may be seen as a collection of connecting objects of a typical size, and then we will argue that for such systems the upper fractal dimension should be $D \approx 2$. We will neglect the variation in the circle size, but a similar argument applies even in the presence of this variation. 

If all the circles have the same size, the void model may be mapped onto a random network of nodes and bonds in the following simple way \cite{kerstein1983equivalence}. We first segment the entire plane into regions, such that all of the points within a given region are closest to one of the circle centers (Figure \ref{fig:6}a). The regions obtained in this way are polygons (known as the Voronoi polygons), and the procedure of segmenting the plane is known as the Voronoi tessellation. Boundaries of these polygons define a network of nodes (the corners of the polygons) and bonds (sides of the polygons). We consider two nodes to be connected if a bond between them does not pass through any of the circles  (Figure \ref{fig:6}b). It was shown \cite{kerstein1983equivalence} that if the circles have a constant radius, nodes that are connected are located within connected voids. This establishes a mapping from the void model to the network. We can then segment each void into pieces and assign each piece to a bond of the network (Figure \ref{fig:6}c). In this way, each bond carries some fraction of the total area and perimeter of the void. Although there is significant variation in how much area and perimeter each individual bond contributes, there is a typical scale above which bond contributions generally do not exist. Therefore, we can imagine that for large enough voids, these variations will average out and each bond will contribute some average amount to the total area and perimeter of the void (Figure \ref{fig:6}d). Variation in the area and perimeter of such large voids will be mainly due to differing numbers of bonds, rather than variation in contributions from individual bonds. Therefore, both the area and perimeter of the void will be proportional to $N$, the number of bonds in a void, $P \propto N$ and $A \propto N$, implying $P \propto A$ and a dimension of $D = 2$. For small voids consisting of just a single bond, area and perimeter will vary due to variation in exact placement of the surrounding circles, and will therefore have a dimension generally less than 2. The beginning of the fractal transition will occur roughly at an area where a two-bond void is as likely as a single-bond void of the same size. The fractal transition will end at a scale where there are enough bonds so that variations due to individual bonds become negligible.

\section{Ponds before drainage}

\begin{figure}
  \centering
   \includegraphics[width=1\textwidth]{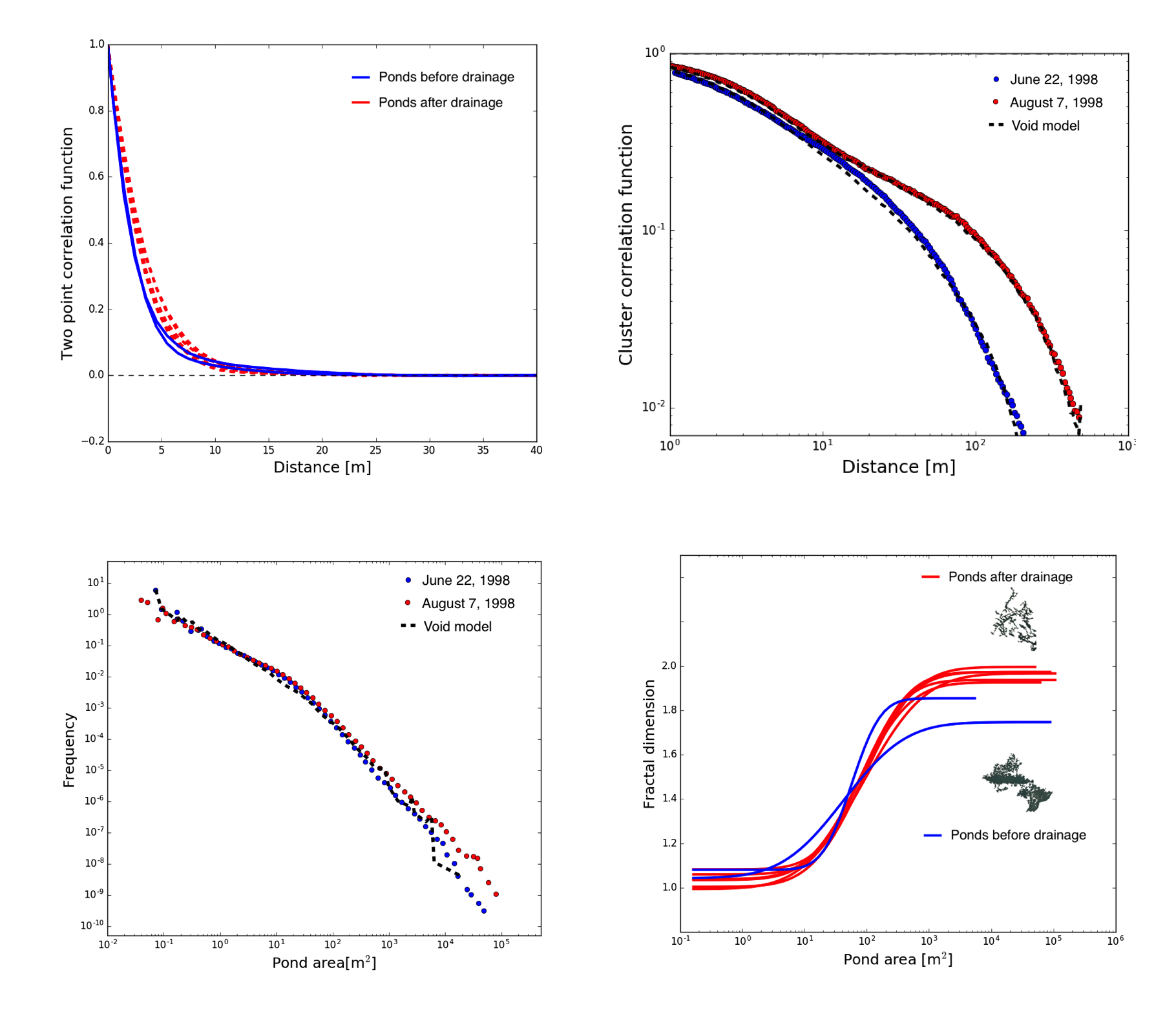}
   \linespread{1.}\selectfont{}
  \caption{Geometric statistics of ponds before drainage: a) A comparison between a two-point correlation function for ponds before drainage (blue lines) and ponds after drainage (red dashed lines). b) A comparison between a cluster correlation function for June 22 (blue circles), August 7 (red circles), and a void model (black dashed lines). c) Pond size distribution for June 22 (blue circles), August 7 (red circles), and a void model (black dashed line). d) Fractal dimension curves for ponds before drainage (blue lines) and after drainage (red lines). Lines end at the maximum pond size found at that date. Examples of a drained and undrained pond are also shown.}
  \label{fig:7}
 \end{figure}
 
We have excluded ponds before drainage from our analysis. This is partly because we do not have reliable data on them: there are only two dates during the SHEBA mission that show ponding before complete drainage, June 15 and June 22. June 15 is the very beginning of the melt season showing only minor pond coverage, while on June 22 only some of the ponds were undrained, making the data inconclusive. Nevertheless, we can proceed to calculate the geometric properties for these dates as well. The two point correlation function shows that both of these dates have roughly the same scale as ponds after drainage, consistent with pre-melt snow dunes setting the pond scale (Figure \ref{fig:7}a). The cluster correlation function and the pond size distribution for ponds on June 15th depend on the threshold parameter, $p_t$, used by the machine learning algorithm to classify the melt ponds, described in section S1. For this reason, these quantities can only be reliably calculated for June 22 (Figures \ref{fig:7}b and \ref{fig:7}c). Both the correlation function and the size distribution for June 22nd can be fit using $\rho = 0.35$, slightly higher than for August 7, and above the percolation threshold. Finally, we find the fractal dimension, $D$, as a function of pond size (Figure \ref{fig:7}d). We do this for both June 15 and June 22, although results for June 15 are inconclusive since there are not many large ponds, so the estimate for the upper fractal dimension has a large error. The upper fractal dimension for both pre-drainage dates is below $D = 2$, in contradiction with the void model. This suggests a qualitative change in the pond morphology before and after drainage. Because of this mismatch and a lack of reliable data, we chose not to apply the void model to pre-drainage ponds.

\section{Many models of melt pond geometry}

\begin{figure}
  \centering
   \includegraphics[width=1\textwidth]{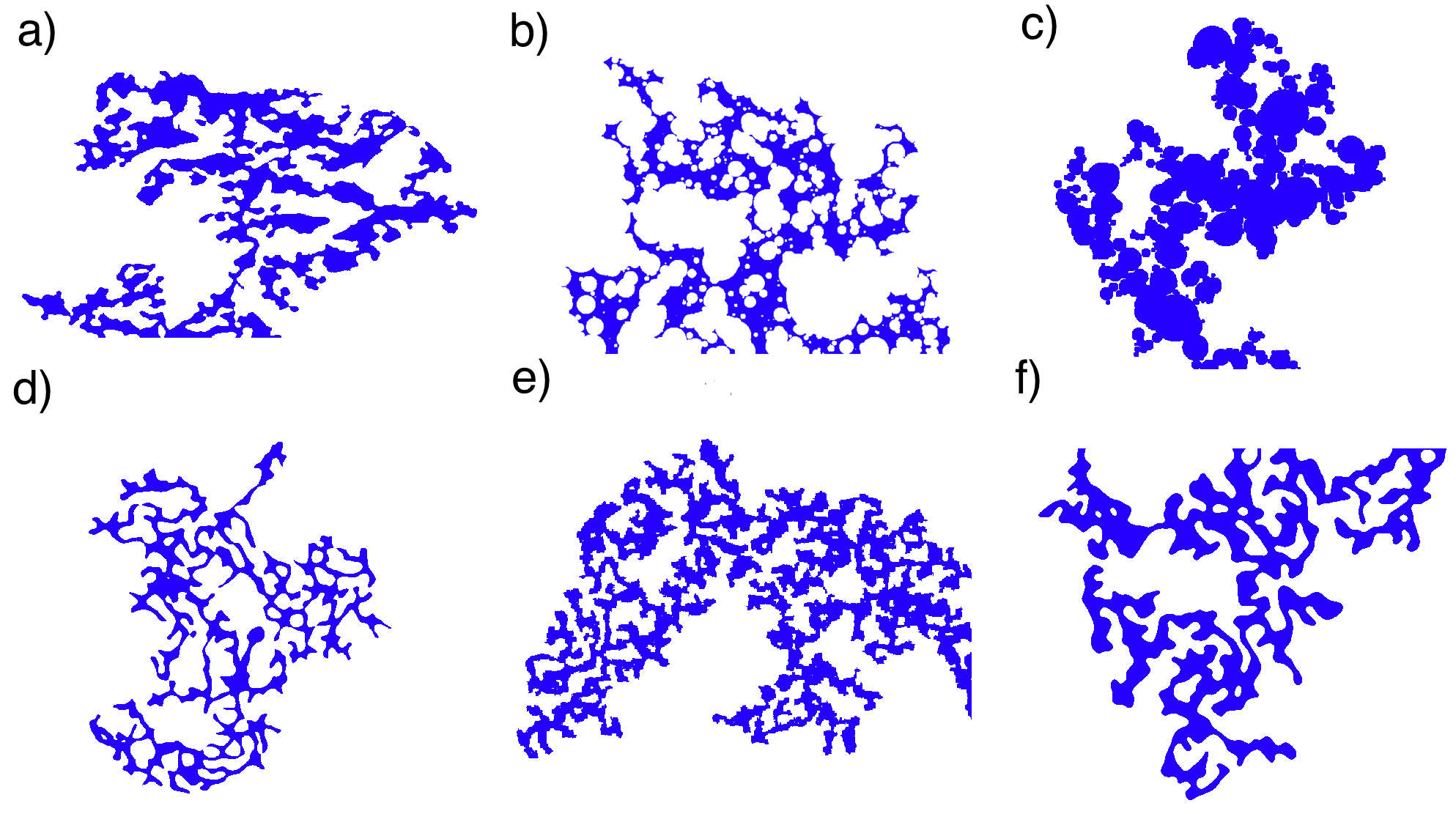}
   \linespread{1.}\selectfont{}
  \caption{Examples of large ponds in different models. a) A real melt pond. b) A void model. c) A circle model. d) A random topography model. e) A quenched Ising model. f) A model of coarsening due to repeated dilations of a binary image with random initial conditions.}
  \label{fig:8}
 \end{figure}

Many models other than the void model are capable of reproducing the geometric features we studied in this paper. These models include a model of overlapping circles, a model of random topography \cite{bowenmodeling}, and several models with coarsening dynamics such as the quenched Ising model \cite{sicilia2007domain,ma2014ising}. We show examples of large melt ponds derived from these models in Figure \ref{fig:8}. All of these models share a common key feature - they represent melt ponds as objects of a typical scale connecting randomly. Any model with such a feature should be able to reproduce the correlation functions, the fractal transition and the size distribution of melt ponds. These models, however, cannot necessarily reproduce all of these pond geometric properties without separately tuning parameters each time. For example, in the complement of the void model, a model of overlapping circles, matching the correlation functions does not yield a correct scale for the fractal transition. The void model is special in that it can robustly match so many pond features with only two parameters that can be independently determined from the data. 

In this paper we only considered the geometry of the melt ponds without addressing their dynamical evolution. Coarsening models include dynamics, and are, therefore, beyond the scope of the current investigation. We note, however, that many of these models (e.g. the quenched Ising model) have the property that, starting from random initial conditions, the connected clusters after a certain period of time are organized near the percolation threshold \cite{sicilia2007domain}. This occurs due to an intrinsic symmetry present in these models between e.g. up-spins and down-spins. Although still far from obvious, this may suggest that some form of coarsening is taking place in melt ponds.

\bibliographystyle{ieeetr}
\bibliography{mybibliog.bib}